\documentclass{emulateapj}

\begin{document}

\title{Origin of a bottom-heavy stellar initial mass function in elliptical
galaxies}

\author{Kenji Bekki} 
\affil{
ICRAR,
M468,
The University of Western Australia
35 Stirling Highway, Crawley
Western Australia, 6009, Australia
}

\begin{abstract}

We investigate the origin of a bottom-heavy stellar initial mass function (IMF) recently observed
in elliptical galaxies by using chemical evolution models  with
a non-universal IMF. We adopt the variable Kroupa IMF with the three slopes 
($\alpha_1$, $\alpha_2$, and $\alpha_3$) dependent on metallicities ([Fe/H])
and densities ($\rho_{\rm g}$)  of
star-forming gas clouds and thereby search for the best IMF model that can reproduce
(i) the observed steep IMF slope ($\alpha_2 \sim 3$, i.e., bottom-heavy) for low stellar masses 
($m\le 1 M_{\odot}$) and (ii) the correlation of $\alpha_2$ with chemical properties
of elliptical galaxies in a self-consistent manner.
We find that if the IMF slope $\alpha_2$ depends both on
[Fe/H] and $\rho_{\rm g}$, then elliptical galaxies with higher [Mg/Fe]
can have steeper $\alpha_2$ ($\sim 3$) in our models.
We also find that 
the observed positive correlation of  stellar mass-to-light ratios ($M/L$)
with [Mg/Fe] in elliptical galaxies
can be quantitatively reproduced in our models with 
$\alpha_2 \propto {\rm \beta[Fe/H]}+\gamma \log \rho_{\rm g}$, where $\beta \sim 0.5$ and $\gamma \sim 2$.
We  discuss whether the IMF slopes for low-mass ($\alpha_2$) and high-mass stars ($\alpha_3$)
need to vary independently from each other to explain a number of IMF-related  observational
results self-consistently.
We also briefly discuss why $\alpha_2$ depends differently on [Fe/H] in
dwarf and giant elliptical galaxies.

\end{abstract}

\section{Introduction}

The stellar initial mass function (IMF) is a principal parameter 
for formation and evolution  of star clusters and galaxies.
Therefore, it has long been discussed observationally and theoretically
whether 
and how the IMF  could vary  with physical conditions of star-forming clouds in galaxies
(e.g., Larson 1998;  Chabrier 2003; Elmegreen 2007; Bastian et al. 2010; Kroupa 2012).
One of important recent observational discoveries regarding the possible IMF variation in galaxies
is that  the IMF in  massive elliptical galaxies could be 
bottom-heavy for low-mass stars  (e.g., van Dokkum \& Conroy 2010; Conroy \& van Dokkum 2012, CV12;
Smith et al. 2012; Spiniello et al. 2012; Ferreras et al. 2013). 
Recent observational studies of early-type galaxies have also revealed a possible
correlation between the IMF slope and galaxy properties such as velocity dispersions
and chemical abundances (e.g., Cenarro et al. 2003; Cappellari et al. 2012;
CV12; Ferreras et al. 2012). 
It is, however, theoretically unclear what physical mechanisms
are behind the observed correlations between the IMF and physical  properties
of early-type galaxies.

CV12 have  investigated the spectral absorption lines
in early-type galaxies in order to provide strong constraints on IMFs of the galaxies by using
their updated  population synthesis model.
 They have found that the IMF for low-mass stars 
becomes increasingly bottom-heavy with increasing velocity dispersions ($\sigma$) and 
[Mg/Fe] in the 38 early-type galaxies.
Although they have found no strong correlations of the IMF with total metallicity ([Z/H]),
it could be possible that there exists a weak/marginal IMF-metallicity correlation.
These results lead the authors to suggest that total metallicity is not a key factor which determines
the IMF slope for low-mass stars.
CV12 have also derived the particular three-part (Kroupa) IMF that can best match the observed spectral indices
and thereby inferred $M/L$ (i.e., they did not directly measure $M/L$).
CV12 have shown a strong correlation of $K$-band mass-to-light
ratios ($M/L_{\rm K}$) normalized to the MW value (($M/L_{\rm K})_{\rm MW}$)
with [Mg/Fe] and briefly discussed the origin of the correlation. 
The origin of the observed $M/L_{\rm K}-{\rm [Mg/Fe]}$ correlation 
has not been extensively investigated by theoretical studies of elliptical galaxy formation.

Narayanan \& Dav\'e (2012) have adopted a broken power-law IMF with
fixed slopes yet variable break-mass depending on star formation rates
and thereby   investigated the IMF evolution of elliptical galaxies
in cosmological simulations.
They have  found that $M/L_{\rm K}$ can be larger for more massive elliptical galaxies with higher
central velocity dispersions ($\sigma$).
Although the simulated $M/L_{\rm K}-\sigma$ correlation derived by 
Narayanan \& Dav\'e (2012)  is consistent qualitatively
with the observed one by CV12,  the simulated slope 
is too shallow to be consistent quantitatively with
the observed one. 
Furthermore their IMF model  with a fixed slope and no dependence of the slope on [Fe/H]
appears to be inconsistent with recent observational results which have shown different IMF slopes
in different galaxies and a dependence of the IMF slope on [Fe/H] (e.g., Geha et al. 2013).
This apparent inconsistency suggests that we need to search for a better IMF model 
that can  explain both (i) the observed bottom-heavy IMF of elliptical galaxies
and (ii) the observed dependences of IMF slopes on physical properties 
(e.g., [Fe/H] and [Mg/Fe]) of galaxies.
Recently Weidner et al. (2013) have pointed out that a time-independent bottom-heavy IMF 
can not explain the observed metallicities of elliptical galaxies and suggested
a two-stage formation scenario.

Marks et al. (2012, M12) have recently proposed a variable Kroupa IMF model
with the three IMF slopes, $\alpha_1$ (for $0.08 \le m/{\rm M}_{\odot} < 0.5$),
$\alpha_2$ ($0.5 \le m/{\rm M}_{\odot} < 1$), and 
$\alpha_3$ ($1 \le m/{\rm M}_{\odot}$).
In their model, $\alpha_1$ and $\alpha_2$ depend on [Fe/H] whereas $\alpha_3$ 
depends  on [Fe/H] and gas densities ($\rho_{\rm g}$)
of star-forming gas clouds.
Their model is promising, firstly because their model
is derived from a detailed comparison between theoretical  and observational results of globular
cluster properties, and secondly because the model can naturally explain
recent observational results on the positive correlation of $\alpha_2$ with [Fe/H] in galaxies
with a wide range of velocity dispersions and metallicities (Geha et al. 2013).
Furthermore,  recent numerical simulations with the variable Kroupa IMF model by M12
have shown that the observed correlation between star formation rate densities and 
the slope of the high-mass end of the IMF ($\alpha_3$)
can be naturally reproduced (Bekki \& Meurer 2013).

However, it is clear that the proposed IMF with a variable slope $\alpha_2$ by M12 
(i.e., $\alpha_2=2.3+0.5{\rm [Fe/H]}$) can not simply explain the observed
strong $\alpha_2-{\rm [Mg/Fe]}$ correlation yet no/little  correlation
between $\alpha$ and metallicities ($Z$ and [Fe/H]) in elliptical galaxies
($\alpha_1$ is assumed to be $\alpha_2-1$ in the present study so that $\alpha_2$ can be
a sole  key parameter for the IMF of low-mass stars in galaxies).
Furthermore,  observational support for the proposed dependence of $\alpha_2$ on [Fe/H]
is significantly  weaker in comparison with $\alpha_3$ in M12.
These facts  imply  that the variable IMF model by M12 needs to be modified significantly
by considering possible dependences of $\alpha_2$ on other physical properties of star-forming
gas clouds, such as gas densities $\rho_{\rm g}$, temperature ($T_{\rm g}$), and 
pressure ($P_{\rm g}$).
Thus it  is particularly important for theoretical studies to investigate how $\alpha_2$ 
in the variable Kroupa IMF needs to  depend
on physical properties
of star-forming clouds so that the observed bottom-heavy IMF and its correlation with
galaxy properties can be self-consistently explained.

The purpose of this paper is to investigate how $\alpha_2$ should depend on 
physical properties of star-forming gas clouds  within galaxies so that the observed
positive correlation between $\alpha_2$ and [Mg/Fe] in elliptical galaxies
can be quantitatively reproduced.
We adopt one-zone chemical 
evolution models of elliptical galaxy formation with a more generalized version
of the variable Kroupa IMF in M12 and thereby
search for the best IMF model that can explain the observed $\alpha_2$-[Mg/Fe] correlation.
We compare the spectroscopically inferred $K$-band $M/L$ (normalized to the MW value)
with the simulated one in order to derive  the best functional form of $\alpha_2$
($=f({\rm [Fe/H]},\rho_{\rm g})$).
In the present study, we consider that observational results
on the $\alpha_2-{\rm [Mg/Fe]}$ (or $M/L - {\rm [Mg/Fe]}$)
correlation  by CV12, which do not show strong
correlations between $\alpha_2$ and metallicities, can be used for determining the best 
variable Kroupa IMF model. However,  Cenarro et al. (2003) reported 
a correlation between the IMF slope and metallicities ([Fe/H]) in elliptical galaxies.
If we use the results by Cenarro et al. (2003) as a constraint on the functional form of
$\alpha_2$, then the best IMF model would be quite different from the one that we can determine
by using the $\alpha_2-{\rm [Mg/Fe]}$ correlation as a constraint.
Therefore, it should be noted that the choice of the best variable IMF model can depend on
which observational results are used as a constraint on the functional form of the IMF.

The layout of this paper is as follows.
In \S 2, we describe our new one-zone chemical evolution
models with a variable IMF model.
In \S 3, we present the results of the time evolution
of $\alpha_2$, [Fe/H], and [Mg/Fe]  for models with different parameters.
In this section, we show the best variable Kroupa IMF model that can reproduce
observations by CV12. 
In \S 4, we discuss IMF-related observational results  that appear to be inconsistent with
a bottom-heavy IMF in elliptical galaxies.
The conclusions of the present study are given in \S 5.
We mainly focus on correlations between IMF slopes and chemical properties of galaxies,
and accordingly do not discuss recent observational
results on the mass-to-light ratios of early-type galaxies
which suggest a non-universal IMF  (e.g., Treu et al. 2010)
in the present study.

\epsscale{1.0}
\begin{figure*}
\plotone{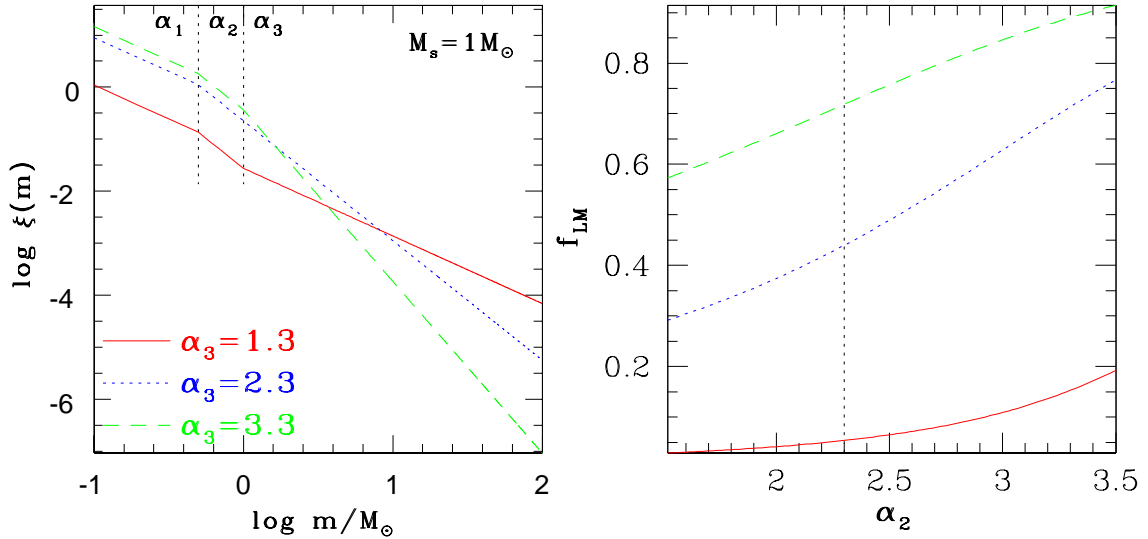}
\figcaption{
Three different variable Kroupa IMF models with $\alpha_3=1.3$ (red, solid), 
2.3 (blue, dotted), 
and 3.3 (green, dashed) for fixed $\alpha_1$ and $\alpha_2$ (left) and the mass fraction
of low-mass stars with $m \le 1M_{\odot}$  ($f_{\rm LM}$)
as a function of $\alpha_2$
for the three IMF models (right).
The total mass of stars ($M_{\rm s}$) is normalized to $1M_{\odot}$ in these models 
and canonical values of 1.3 and 2.3 are adopted for $\alpha_1$ and $\alpha_2$, respectively,
in the left panel.
The dotted line in the right panel indicates the canonical $\alpha_2$.
\label{fig-1}}
\end{figure*}

\section{The  model}

\subsection{Outline}

\subsubsection{Comparing the spectroscopically implied M/L with  the modeled ones}

The main purpose of this study is to compare the observed and modeled $M/L-{\rm [Mg/Fe]}$
relations.
First of all, it should be noted that CV12 derived $M/L$ of elliptical galaxies from
the gravity-sensitive spectral features by adopting a variable Kroupa IMF (with fixed $\alpha_3$ yet
variable $\alpha_1$ and $\alpha_2$). Therefore, the $M/L$ in CV12 is not a direct measurement of  $M/L$
and it can  depend on the modeling of IMF. We however compare the observationally inferred $M/L$ 
by CV12 with the modeled ones
in order to derive physical meanings of the observed $M/L-{\rm [Mg/Fe]}$ relation. Also, as described later,
a number of assumptions are made in deriving $M/L$ of elliptical galaxies from chemical evolution models. 
Thus, there are some limitations both in inferring $M/L$ from the observed spectral features in CV12
and in modeling  $M/L$ in the present study.

\subsubsection{Two approximations in M/L modeling}

We need to derive both [Mg/Fe] and $M/L$ by using the present one-zone
chemical evolution model and a publicly available stellar population synthesis code. 
Although it is straightforward to calculate [Mg/Fe] in the present one-zone model,
we need to take the following steps to estimate $M/L$. First we derive $\alpha_2$
(and $\alpha_1=\alpha_2-1$) of a galaxy by using the adopted variable Kroupa IMF that depends
on gas densities and [Fe/H] in one-zone chemical evolution
models.   Then we estimate the mean  $M/L$
of old stellar populations of the galaxy  from the derived $\alpha_2$ (that is integrated
over all time steps). It  would be ideal that we adopt a fully self-consistent one-zone model
with chemical yields depending on $\alpha_1$, $\alpha_2$, and $\alpha_3$ 
at each time step and
use a stellar population synthesis code to calculate $M/L$ of old stellar populations
with different Kroupa IMF slopes. 
However, 
most of stellar population synthesis codes are for a fixed IMF 
(but see Conroy et al. 2009 for a new  model with a variable Kroupa IMF)
and we have not yet 
developed one-zone chemical evolution models with variable Kroupa IMFs.
Therefore, we have to adopt the following two
approximations as a compromise.

One is that chemical yields do not depend on $\alpha_2$ ($\alpha_1$)  in the present
one-zone models in which $\alpha_2$ is assumed to vary with time.  
This approximation can be justified as follows.
Chemical yields from supernovae and AGB stars
depend much more strongly on $\alpha_3$ than on $\alpha_1$ and $\alpha_2$.
Therefore, the inclusion of time-varying $\alpha_1$ and $\alpha_2$ 
in one-zone chemical evolution models would not
change significantly the present results. In order to demonstrate this
point,  we have investigated [Mg/Fe]$-$[Fe/H] relations of
elliptical galaxies
with different $\alpha_1$ and $\alpha_2$ (but fixed $\alpha_3=2.3$,
like the Salpeter IMF) in the variable Kroupa IMF
and the results and their discussion are shown in
Appendix A. Clearly, the final [Mg/Fe] does not depend strongly on $\alpha_1$ and
$\alpha_2$ (but it depends more strongly on star formation time scales of
galaxies).
 Thus the adopted approximation can be regarded as good enough to
discuss the final  [Mg/Fe] of galaxies.

The other is that $M/L$ predicted for  a  single-power-law IMF 
rather than $M/L$ for a (variable/fixed) Kroupa IMF is used for estimating $M/L$ of a galaxy.
The adoption of this approximation means that the present model is not
self-consistent (i.e., using a stellar population synthesis code 
based on a variable single-power-law
IMF for chemical evolution models with a variable Kroupa IMF).  
However,
this is the best that we can do,
because  most of  publicly available stellar population
synthesis codes are  for a fixed IMF and we have used them so far.
In order to demonstrate whether this
approximation is good enough to discuss $M/L$, 
we have investigated  possible $M/L$ difference in  single-power-law
and variable Kroupa IMFs with different $\alpha_2$ and 
the results are shown  in Appendix B.

It is clear in Appendix B  that (i) the absolute values of $M/L$ 
can be slightly different between single-power-law ($\alpha_1=\alpha_2$) and variable Kroupa 
($\alpha_1=\alpha_2-1$) IMFs for a given $\alpha_2$ ($=\alpha_3$)
and (ii) the $M/L$ difference does not depend  strongly  on $\alpha_2$. 
These mean that  single-power-law IMFs are  highly likely to  overestimate
$M/L$ by  a similar amount in comparison with the variable Kroupa IMFs (with the same $\alpha_3$)
for a wide range of $\alpha_2$ ($=\alpha_3$).
These accordingly demonstrate  that the slope of the 
$M/L-{\rm [Mg/Fe]}$ relation modeled in the present study
(in which a stellar population synthesis code for a variable single-power-law IMF 
is adopted)  can be very close to the true one
that is derived self-consistently by using a stellar population synthesis code
for a variable Kroupa IMF.   Therefore,  a comparison between the observed and modeled slopes
can be regarded as reasonable.
It should be noted, however, that $\alpha_1$ and $\alpha_2$ in  the IMF adopted by CV12 
can vary independently ($0 \le \alpha_1, \alpha_2 \le 3$) and therefore their IMF is different
from ours in which $\alpha_1=\alpha_2-1$.

Thus, the present model  is not fully self-consistent in terms of the derivation of $M/L$
from the outputs of one-zone chemical evolution models  owing to
the adoption of the above two approximations.
However, as long as we discuss the slope of the observed
$M/L-{\rm [Mg/Fe]}$ relation, the present model greatly helps us to extract some
important physical meanings of the observed $M/L-{\rm [Mg/Fe]}$ relation.
 We will be able to  more properly estimate $M/L$ in our future studies by
adopting the latest  stellar population synthesis code for any combinations of $\alpha_1$, $\alpha_2$,
and $\alpha_3$ of a variable  Kroupa IMF  (e.g., Conroy et al. 2009).
We consider that the main conclusion of this paper  
(i.e., $\alpha_2$ should be proportional to  $\sim 0.5 {\rm [Fe/H]}+2\log \rho_{\rm g}$)
will not change significantly in our future
better models, because the conclusion is derived from a comparison between the observed
and modeled slopes of the $M/L-{\rm [Mg/Fe]}$ relation (not from a comparison between
the observed and simulated absolute values of $M/L$ themselves).

\begin{deluxetable}{ccccc}
\footnotesize
\tablecaption{Model parameters for the 20 representative one-zone chemical evolution models
\label{tbl-1}}
\tablewidth{0pt}
\tablehead{
\colhead{  Model } &
\colhead{  $t_{\rm a}$  \tablenotemark{a} } &
\colhead{  $t_{\rm trun}/t_{\rm a}$  \tablenotemark{b} } &
\colhead{  $C_{\rm sf}$ \tablenotemark{c} }   &
\colhead{  IMF type \tablenotemark{d} }  } 
\startdata
M1 & 0.5 & 2 & 0.4 & Salpeter\\
M2 & 1.0 & 2 & 0.4 & Salpeter \\
M3 & 2.0 & 2 & 0.4 & Salpeter \\
M4 & 3.0 & 2 & 0.4 & Salpeter  \\
M5 & 4.0 & 2 & 0.4  & Salpeter\\
M6 & 0.5 & 1 & 0.4 & Salpeter\\
M7 & 1.0 & 1 & 0.4 & Salpeter \\
M8 & 2.0 & 1 & 0.4 & Salpeter\\
M9 & 3.0 & 1 & 0.4 & Salpeter\\
M10 & 4.0 & 1 & 0.4 & Salpeter \\
M11 & 0.5 & 2 & 0.2 & Salpeter \\
M12 & 1.0 & 2 & 0.2 & Salpeter \\
M13 & 2.0 & 2 & 0.2 & Salpeter \\
M14 & 3.0 & 2 & 0.2 & Salpeter\\
M15 & 4.0 & 2 & 0.2 & Salpeter \\
M16 & 0.5 & 2 & 0.8 & Salpeter \\
M17 & 1.0 & 2 & 0.8 & Salpeter \\
M18 & 2.0 & 2 & 0.8 & Salpeter \\
M19 & 3.0 & 2 & 0.8 & Salpeter\\
M20 & 4.0 & 2 & 0.8 & Salpeter \\
M21 & 0.5 & 2 & 0.4 & Kroupa, $\alpha_1=1.3$, $\alpha_2=2.3$ \\
M22 & 0.5 & 2 & 0.4 & Kroupa, $\alpha_1=2.3$, $\alpha_2=2.3$ \\
M23 & 0.5 & 2 & 0.4 & Kroupa, $\alpha_1=2.3$, $\alpha_2=3.3$ \\
M24 & 0.5 & 2 & 0.4 & Kroupa, $\alpha_1=0.3$, $\alpha_2=1.3$ \\
\enddata
\tablenotetext{a}{The gas accretion timescale in units 
of Gyr.}
\tablenotetext{b}{The ratio of the SF truncation timescale  ($t_{\rm trun}$)
to the gas accretion timescale $t_{\rm a}$.}
\tablenotetext{c}{The dimensionless parameter that controls SF rates. The larger $C_{\rm sf}$ is 
in a galaxy of  a model,
the higher the SFR is.}
\tablenotetext{d}{For the single-power-law Salpeter IMF (M1$-$M20), 
$\alpha=2.35$ (i.e., $\alpha_1=\alpha_2=\alpha_3=2.35$) 
is adopted for all models.  For the Kroupa IMF (M21$-$M24), 
different $\alpha_1$ and $\alpha_2$ are adopted, but $\alpha_3$ is fixed at 2.3.}
\end{deluxetable}

\subsection{The variable  Kroupa IMF}

We consider that the three slopes in the variable Kroupa IMF can vary according to the 
physical conditions  of star-forming regions in the present study (this variable Kroupa
IMF is illustrated in Figure 1).
In the original M12's IMF,
the low-mass end of the variable Kroupa IMF
($\alpha_1$ for $0.08 \le m/{\rm M}_{\odot} < 0.5$) depends solely on [Fe/H]
as follows
\begin{equation}
\alpha_1=1.3+0.5\times {\rm [Fe/H]}.
\end{equation}
The value of $\alpha_2$ for $0.5 \le m/{\rm M}_{\odot} < 1$
is also determined
solely by [Fe/H];
\begin{equation}
\alpha_2=2.3+0.5\times {\rm [Fe/H]}.
\end{equation}
The high-mass end of the variable Kroupa IMF 
$\alpha_3$ for
$1 \le m/{\rm M}_{\odot} \le 100$
is described as follows;
\begin{equation}
\alpha_3 = 0.0572 \times {\rm [Fe/H]} -0.4072 \times \log(
\frac{ \rho_{\rm cl} }{ 10^6 M_{\odot} {\rm pc}^{-3} }) +1.9283,
\end{equation}
where $\rho_{\rm cl}$ is the density of a rather high-density  gaseous
core where star formation can occur.
This equation holds for $x_{\rm th} \ge -0.87$, where
$x_{\rm th}=-0.1405{\rm [Fe/H]}+\log(
\frac{ \rho_{\rm cl} }{ 10^6 M_{\odot} {\rm pc}^{-3} })$,
and $\alpha_{3}=2.3$ for $x_{\rm th} <-0.87$ (M12).  
Although the two coefficients in the equation (3) are precisely described,
they are simply the best-fit parameter values of their IMF that can explain observations
(i.e., Observations can not determine the coefficients with  such a high precision).
In the present study, 
unlike in  M12,
it is assumed that
$\alpha_3$ does not  vary with densities and [Fe/H].

As briefly discussed in \S 1,
the observed correlation between the IMF slope and chemical abundances in elliptical galaxies
(CV12) can not be simply explained by the above equation (2). We therefore adopt the
following  more generalized
version of the variable IMF for $\alpha_2$: 
\begin{equation}
\alpha_2=\alpha_{2,s}+\beta \times {\rm [Fe/H]} + \gamma \times \log \rho_{\rm g},
\end{equation}
where $\alpha_{\rm 2,s}$ is the value for
the solar neighborhood and $\rho_{\rm g}$ is the gas density
of a star-forming gas cloud. 
This $\rho_{\rm g}$ is the mean density of a star-forming cloud
and thus different from the density of a molecular core ($\rho_{\rm cl}$)
in the equation (3).
This functional form ($\alpha_2=f({\rm [Fe/H]},\rho_{\rm g})$)
needs to ensure that $\alpha_2 \sim 2.3$ at the solar neighborhood. Therefore, the above
equation is modified as follows.
\begin{equation}
\alpha_2=2.3+\beta \times {\rm [Fe/H]} + \gamma \times \log 
\frac { \rho_{\rm g} + \rho_{\rm th} } { \rho_{\rm s}+\rho_{\rm th} },
\end{equation}
where $\rho_{\rm s}$ is the typical gas density for star-forming gas clouds at the solar neighborhood
and $\rho_{\rm th}$ is introduced so that $\alpha_2$ can not be too small for low-density star-forming
regions. This $\rho_{\rm th}$ can correspond to a threshold gas density beyond
which star formation can occur. In the present study, $\alpha_1$ is assumed to be $\alpha_2-1$
throughout this paper,
 though in reality  $\alpha_1$ could in principle  vary independently from $\alpha_2$.
 We consider that gas density is a more fundamental parameter for $\alpha_2$ than SFR,  because
SFR can depend not only on gas density  but also on other gas properties (e.g., molecular content
and dynamical time scale). Also, the adopted relation between the IMF slope and  gas density
for $\alpha_2$ 
is more consistent with that for $\alpha_3$ in which gas density rather than SFR is  a key parameter.

Our one-zone chemical evolution models (later described) can output [Fe/H] and $\rho_{\rm g}$
so that we can investigate the time evolution of  $\alpha_2$ for a given  $\beta$ and $\gamma$
by using the equation (5).
Although we have investigated the models with $\beta=0$ and 0.5,
we show the results of the models with $\beta=0.5$. This is firstly because
$\beta=0.5$ is consistent with 
recent observations on the dependence of $\alpha_2$ on [Fe/H] (Geha et al. 2013),
and secondly because the models with $\beta=0$ can not explain the observed $M/L-{\rm [Mg/Fe]}$
relation.
We investigate models with different $\gamma$ (=0, 0.5, 1.0, 1.5, 2.0, and 2.5)
and thereby try to find models which can reproduce the observed $M/L_{\rm K}-{\rm [Mg/Fe]}$
correlation (which corresponds to a $\alpha_2 - {\rm [Mg/Fe]}$ correlation)
in a quantitative manner.
In the present study,  we use chemical evolution models just for the purpose of finding
the best variable Kroupa IMF with a certain value of $\gamma$.
Accordingly, in computing the predicted [Mg/Fe],  we adopt a standard (yet simple) Salpeter IMF model
($\alpha_1=\alpha_2=\alpha_3=2.35$) in which 
the IMF slope is fixed during chemical evolution of a galaxy. A justification of adopting such a model
is given in Appendix A.

\epsscale{1.0}
\begin{figure*}
\plotone{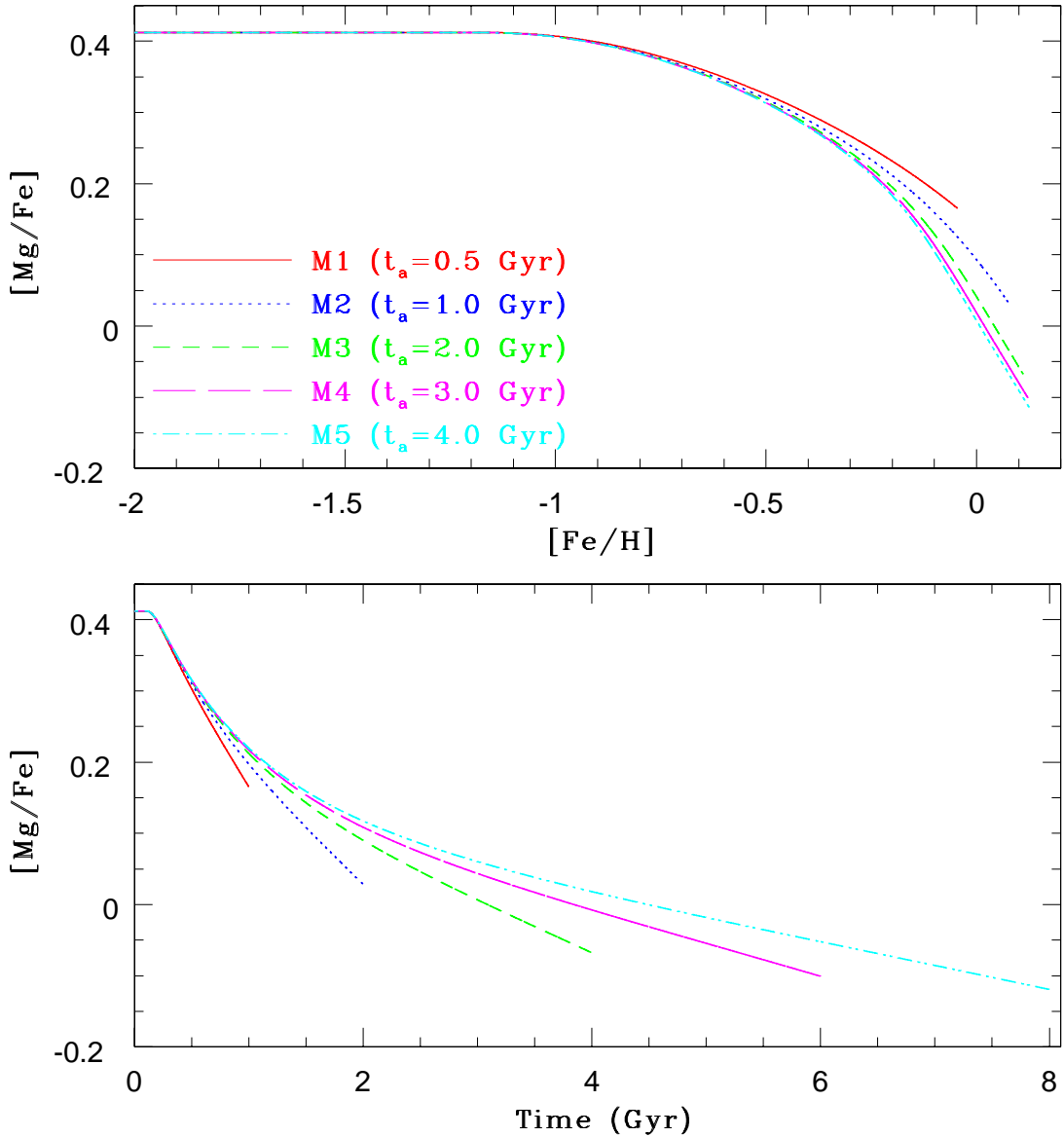}
\figcaption{
The evolution of galaxies on the [Mg/Fe]$-$[Fe/H] plane (upper) and the time evolution
of [Mg/Fe] for the five chemical evolution models, 
M1 with $t_{\rm a}=0.5$ Gyr (red, solid), 
M2 with $t_{\rm a}=1$ Gyr (blue, dotted),
M3 with $t_{\rm a}=2$ Gyr (green, short-dashed),
M4 with $t_{\rm a}=3$ Gyr (magenta, long-dashed), 
and M5 with $t_{\rm a}=4$ Gyr (cyan, dot-dashed).
The star formation is assumed to be truncated at $t=2t_{\rm a}$ in these models.
\label{fig-2}}
\end{figure*}

\subsection{Chemical evolution}

Elliptical galaxies are assumed to form with initial massive starbursts at high redshifts,
as often assumed
in previous chemical evolution models (e.g., Arimoto \& Yoshii 1987; Matteucci et al. 1998;
Pipino \& Matteucci 2004).  
The duration of the initial starbursts is assumed to be different in different models
so that the final elliptical galaxies can have different [Fe/H] and [Mg/Fe] in the present study.
We do not discuss other important aspects of elliptical galaxy formation,
such as the origin of the color-magnitude 
or mass-metallicity relations among elliptical galaxies with different
masses and luminosities.
We adopt one-zone chemical evolution models that are essentially the same
as those adopted in our previous studies on the chemical
evolution of the Large Magellanic Cloud, LMC (Bekki \& Tsujimoto 2012, BT12).
Accordingly, we briefly describe the adopted models
in the present study.

We investigate the time evolution of the gas mass fraction
($f_{\rm g}(t)$), the star formation rate ($\psi(t)$), and
the abundance of the $i$th heavy element ($Z_i(t)$) for a given
accretion rate ($A(t)$), IMF,
and ejection rate of ISM ($w(t)$).
The basic equations for the adopted one-zone chemical evolution models
are described as follows:
\begin{equation}
\frac{df_g}{dt}=-\alpha_{\rm lock}\psi(t)+A(t)-w(t)
\end{equation}
\begin{eqnarray}
\frac{d(Z_if_g)}{dt}=-\alpha_{\rm lock} Z_i(t)\psi(t)+Z_{A,i}(t)A(t)+y_{{\rm II},i}\psi(t) 
\nonumber \\  +y_{{\rm Ia},i}\int^t_0
\psi(t-t_{\rm Ia})g(t_{\rm Ia})dt_{\rm Ia}
\nonumber \\ +\int^t_0
y_{{\rm agb},i}(m_{\rm agb})
\psi(t-t_{\rm agb})h(t_{\rm agb})dt_{\rm agb}
-W_i(t) \ \ ,
\end{eqnarray}
\noindent where $\alpha_{\rm lock}$ is the mass fraction
locked up in dead stellar remnants and long-lived stars,
$y_{\rm {Ia}, i}$, $y_{\rm {II}, i}$, and
$y_{\rm {agb}, i}$ are the
chemical yields for the $i$th element from type II supernovae (SN II),
from SN Ia,
and from AGB stars, respectively,
$Z_{A,i}$ is the abundance of heavy elements  contained in the infalling gas,
and $W_i$ is the wind rate for each element.
The quantities $t_{\rm Ia}$ and $t_{\rm agb}$ represent
the time delay between star formation and SN Ia explosion
and that between star formation and the onset of AGB phase,
respectively.
The terms $g(t_{\rm Ia}$) and $h(t_{\rm agb})$ are the distribution
functions of SNe Ia and AGB stars, respectively.
The term $h(t_{\rm agb})$ controls
how much AGB ejecta can be returned into the ISM per unit mass for a given time
in equation (7).
The total gas masses  ejected from AGB stars depends on the original masses
of the AGB stars (e.g., Weidemann 2000).
Therefore, this term  $h(t_{\rm agb})$ depends on
the adopted IMF and the age--mass relation of the stars.
We adopt the same models for $g(t_{\rm Ia})$ and $h_{\rm agb}$ as those used
in BT12. The wind and ejection rates ($w(t)$ and $W(t)$, respectively)
are  set to be 0 in all models of the present study.
Thus equation (6) describes the time evolution
of the gas due to star formation and  gas accretion.
Equation (7) describes
the time evolution
of the chemical abundances due to chemical enrichment by supernovae and AGB stars.

The star formation rate $\psi(t)$ is assumed to be proportional
to the gas fraction with a constant star formation
coefficient and thus is described as follows:
\begin{equation}
\psi(t)=C_{\rm sf}f_{\rm g}(t).
\end{equation}
This $C_{\rm sf}$ given in dimensionless units can control the strength of a starburst
in each model and its value  is assumed to be different between different models.
The star formation is assumed to be truncated at $t_{\rm trun}$, after which
elliptical galaxies can evolve passively without star formation.
For the accretion rate, we adopt the formula in
which $A(t)=C_{\rm a}\exp(-t/t_{\rm a})$
and $t_{\rm a}$ is a free parameter controlling the time scale of
the gas accretion.   The normalization factor $C_{\rm a}$ is determined such
that the total gas mass accreted onto an elliptical galaxy  can be 1 for a given
$t_{\rm a}$ and $t_{\rm trun}$.  Although we investigated  models with different $t_{\rm a}$,
we show the models with $t_{\rm a}=0.5$, 1, 2, 3, and 4 Gyr.
The initial [Fe/H] of the infalling  gas is set to be $-2$  and
we assume a SN-II like enhanced [$\alpha$/Fe] ratio
(e.g., [Mg/Fe]$\approx 0.4$) for the gas.

We adopt  the nucleosynthesis yields of SNe II and Ia from Tsujimoto et al. (1995)
to deduce $y_{{\rm II},i}$ and $y_{{\rm Ia},i}$ for a given IMF.
We adopt a fixed  Salpeter IMF,
$dN/dm=\xi(m) \propto  m^{-\alpha}$, where $\alpha$ is 
the IMF slope and fixed at 2.35
for the calculation of chemical yields.
The fraction of the stars that eventually
produce SNe Ia for 3--8$M_{\odot}$ has not been observationally determined
and thus is regarded as a free parameter, $f_{\rm b}$. Although we
investigate models with $f_{\rm b}=0.05$, 0.1, and 0.15, we  describe
the models with $f_{\rm b}=0.05$. The present chemical evolution models
of elliptical galaxies with $f_{\rm b}=0.05$ can show the {\rm mean} [Mg/Fe]
as low as $\sim 0.1$ only
after $\sim 6$ Gyr evolution.  If we adopt $f_{\rm b}=0.15$ (as Pipino \& Matteucci 2004 did),
then the mean [Mg/Fe] can more rapidly become as low as 0.1.
We consider that using 20 representative models with $f_{\rm b}=0.05$ is enough
to find the best $\gamma$ for successful reproduction of observations regarding the IMF variation
in elliptical galaxies.
The parameter values of the twenty representative
models (M1$-$M20) and four additional ones (M21$-$M24) with 
the variable Kroupa IMF models for Appendix A  are summarized in Table 1.

\begin{figure*}
\plotone{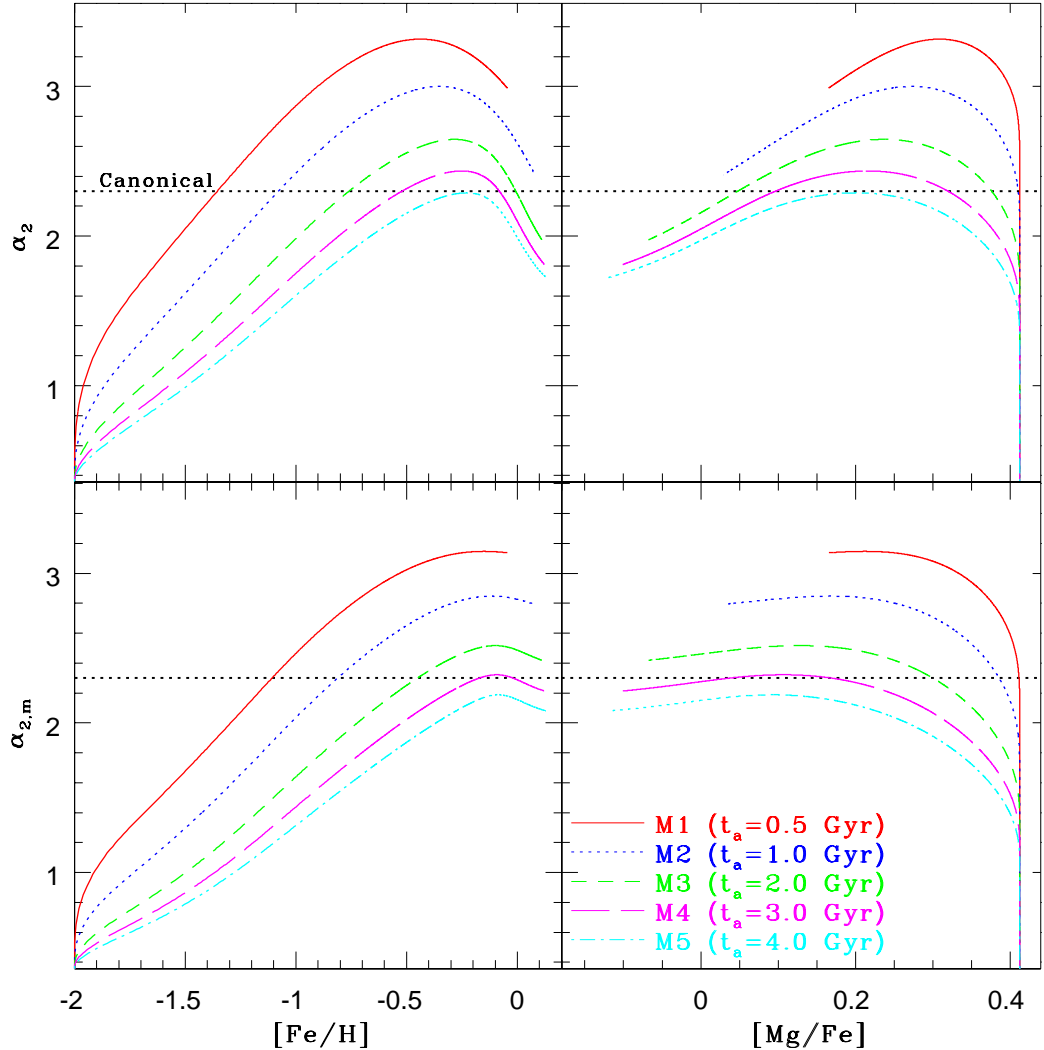}
\figcaption{
The evolution of $\alpha_2$ (upper two) and $\alpha_{\rm 2,m}$ (lower two) 
as a function of [Fe/H] (left) and [Mg/Fe] (right) for the five models,
M1 (red, solid),
M2 (blue, dotted),
M3 (green, short-dashed),
M4 (magenta, long-dashed),
and M5 (green, dot-dashed).
Here $\alpha_2$ is an instantaneous  value at each time  ($t$) whereas $\alpha_{\rm 2,m}$ is
the mean averaged for all star formed by the time $t$.
The dotted lines indicate a canonical value of $\alpha_2$.
\label{fig-3}}
\end{figure*}

\subsection{Derivation of $\alpha_2$ and $M/L$}

In order to estimate $\alpha_2$ by using equations (5), (6), (7), and (8),
the (typical) gas density of a star-forming cloud in a forming elliptical galaxy
at each time step is calculated as follows:
\begin{equation}
\rho_{\rm g}(t) =C_{\rm den} F_{\rm g}(t),
\end{equation}
where $C_{\rm den}$ is a normalization factor for $\rho_{\rm g}$ and 
$F_{\rm g}(t)$ is the ratio of the total gas mass at a time $t$ to the total mass
of gas and stars at $t=t_{\rm trun}$. This $F_{\rm g}$ is calculated from equations (6) and (7).
In the present study, $\rho_{\rm s}$ and $\rho_{\rm th}$
are set to be 1 and 0.5, respectively. Therefore, $C_{\rm den}$ should be $10-20$ to ensure
that a Milky Way-like 
galaxy with $F_{\rm g} = 0.05 \sim 0.1$ can have $\alpha_2=2.3$ in the equation (5).
Although we  investigate the models with $C_{\rm den}=5$, $10$, and $20$,
we show the results of the models with $C_{\rm den}=20$ (i.e., higher density of gas clouds).
This is because the models with $C_{\rm den}=5$ and $10$ do not show high $\alpha_2$ as observed.
Such models with lower $C_{\rm den}$ would be reasonable for disk galaxies with low mean mass
densities.

As mentioned in \S 2.1,
although  we adopt a variable Kroupa IMF in our chemical evolution models, 
we can not calculate
$M/L$ for the modeled galaxies in a fully self-consistent manner, because of the lack of stellar
population synthesis codes for the variable 
Kroupa IMF with different three IMF slopes.
Therefore, as a compromise, we use the code ``MILES'', which is a new population synthesis code
for a variable IMF with a single power-law form and made publicly available by Vazdekis et al. 
(2010). 
The MILES can output $M/L$ for different metallicities and different IMF slopes
(for a single power-law IMF with a slope $\alpha$).
At each time step in a one-zone chemical evolution model,
$\alpha_1$ and $\alpha_2$ can be derived by using the equations (5)
and a relation of $\alpha_1=\alpha_2-1$.
Since the MILES adopts a fixed single lower-law slope
(i.e., $\alpha_1=\alpha_2=\alpha_3=\alpha$), 
we have to use the derived  $\alpha_2$ as 
$\alpha$ in the MILES
and  thereby estimate $M/L$ by 
using  the tabulated values of $M/L$ in the SSP of the MILES.
For example, if $\alpha_2=2.5$ (thus $\alpha_1=1.5$)  at a time step of a model,
then we use the tabulated $M/L$ of $\alpha=2.5$ 
(i.e., $\alpha_1=\alpha_2=\alpha_3=2.5$)
for the age (and metallicity) of a stellar population  formed at
the time step in order to calculate the  $M/L$.
We estimate the mean $M/L$ by using $M/L$ of all stars formed in a model.

It should be noted that CV12 did not measure the masses of elliptical galaxies
but instead compute the expected masses from the IMF required to fit the observed
spectrum of the galaxies by using the observed absorption lines that are sensitive
to stellar gravity. 
Furthermore, the MILES and
an original SSP code are used in this study and CV12, respectively.
Therefore, the $M/L$ normalized by the MW value in CV12
is not exactly the same as $M/L$ calculated in the present study.
The simulated $M/L$ is normalized by $M/L$ for a SSP with a solar metallicity and
an age of 12.6 Gyr so that the simulated $M/L$ range can be similar to the observed
one by CV12. This normalization is done just for convenience.
We consider 
that a comparison between the present results and observational ones  by CV12 enables us
to derive the best variable Kroupa IMF model. 

\begin{figure}
\plotone{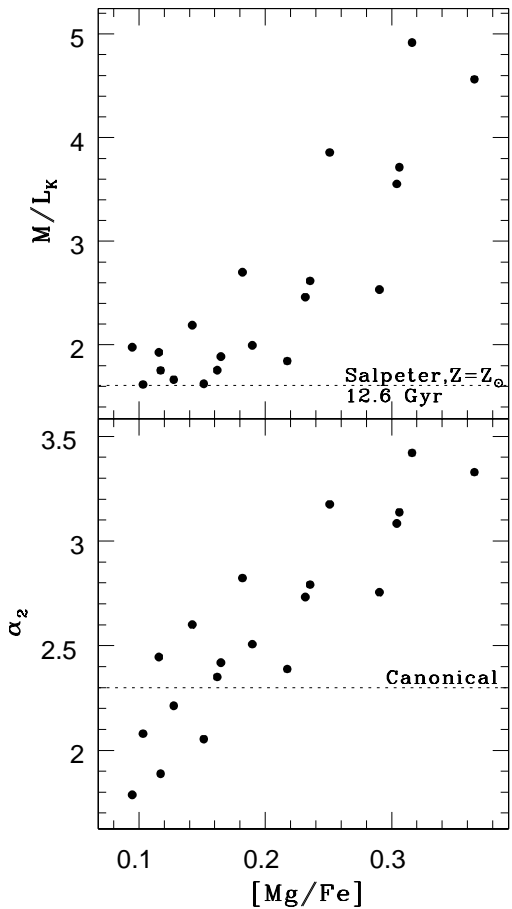}
\figcaption{
The dependences of $M/L_{\rm K}$ (upper) and $\alpha_2$ (lower) on [Mg/Fe] for 20 models
with different $t_{\rm a}$, $t_{\rm trun}$, and $C_{\rm sf}$. The mean $M/L_{\rm K}$ and $\alpha_2$
are estimated from  all stars in each model. The canonical $\alpha_2$ (=2.3) and $M/L_{\rm K}$
for the Salpeter IMF with a solar metallicity and an age of 12.6 Gyr   are shown
by dotted lines.
\label{fig-4}}
\end{figure}

\section{Results}

\subsection{$\alpha_2$ evolution}

Figure 2 shows the time evolution of [Mg/Fe]
and galaxy evolution on the [Mg/Fe]-[Fe/H] plane for the five representative model
with $C_{\rm sf}=0.4$ and different $t_{\rm a}$. Owing to the adoption of the prompt
SNIa model,  [Mg/Fe] can rapidly decrease with time  from the early evolutionary phase ($t<0.2$ Gyr)
for these models. The models with shorter $t_{\rm a}$ thus stronger initial starburst
can have larger final [Mg/Fe] and smaller [Fe/H] in these models. For the adopted $f_{\rm b}=0.05$
in these models,
active star formation needs to continue at least $\sim 6$ Gyr so that galaxies can have 
the {\it mean} [Mg/Fe] (not the instantaneous  one as shown in this figure) as small as 0.1.
Although models with such a relatively long continuous star formation might not be reasonable for 
giant elliptical galaxies,  observations by CV12 showed
that some elliptical galaxies have  [Mg/Fe]$\sim 0.1$.
We therefore consider that  these model (M4 and M5) can represent stellar populations of some
elliptical galaxies with [Mg/Fe]$\sim 0.1$. 

Figure 3 shows the time evolution of galaxies on the $\alpha_2-$[Fe/H],
$\alpha_{\rm 2, m}-$[Fe/H],
$\alpha_2-$[Mg/Fe],
and $\alpha_{\rm 2,m}-$[Mg/Fe] planes for the five models with $\gamma=2.0$.
Here $\alpha_2$ is an instantaneous  value at a  time $t$ whereas $\alpha_{\rm 2, m}$ is the mean 
$\alpha_2$ averaged for
all stars formed by the time  $t$.
The values of [Fe/H] and [Mg/Fe] are those at a given time step (not the average over all stars
formed to that time step).
Since $\alpha_2$ is proportional to $0.5{\rm [Fe/H]} +2\log\rho_{\rm g}$,
$\alpha_2$ can increase with time owing to (i) increasing [Fe/H] and (ii) higher $\rho_{\rm g}$
in these models. After $\alpha_2$ reaches its peak value at a certain [Fe/H],
it start to decrease owing to the lower $\rho_{\rm g}$. In this decreasing phase,
$\alpha_2$ can decrease with decreasing [Mg/Fe]. The models with higher SF can show
both higher final mean $\alpha_2$ ($\alpha_{\rm 2,m}$) and higher [Mg/Fe], which means
that there should be a  positive correlation between [Mg/Fe] and $\alpha_2$.
Clearly, the models M1 and M2 show the final mean $\alpha_2$ significantly larger (i.e., steeper)
than the canonical Salpeter IMF ($\alpha_2=2.3$), which is consistent with observational
results by CV12.

Figure 4 shows the locations of final elliptical 
galaxies on the $M/L_{\rm K}-$[Mg/Fe] and $\alpha_2-$[Mg/Fe] planes
in the 20 models with different $C_{\rm sf}$, $t_{\rm a}$, and $t_{\rm trun}$ for  $\gamma=2.0$. 
It is clear that galaxies with higher [Mg/Fe] can have higher $M/L_{\rm K}$ and larger
$\alpha_2$ in these models. 
Owing to the adopted dependence of $\alpha_2$ both on [Fe/H] and on $\rho_{\rm g}$,
there can be a dispersion in $M/L_{\rm K}$ and $\alpha_2$ for a given [Mg/Fe]. It should be stressed
that the present models can reproduce not only the bottom-heavy IMF ($\alpha_2 \sim  3$) at high
[Mg/Fe] but also an IMF shallower than the Salpeter at low [Mg/Fe]. 
The original variable Kroupa IMF model  for $\alpha_2$ (M12)  depends only on [Fe/H],
and therefore $\alpha_2$ can be only  2.4 at [Fe/H]=0.2 in the model. This value of $\alpha_2$ is 
significantly smaller than the observed bottom-heavy IMF ($\alpha_2 \sim 3$)
for metal-rich giant elliptical galaxies, which means
that the present variable IMF model has an advantage in reproducing the observed large
$\alpha_2$ in giant elliptical galaxies.

\begin{figure*}
\epsscale{1.0}
\plotone{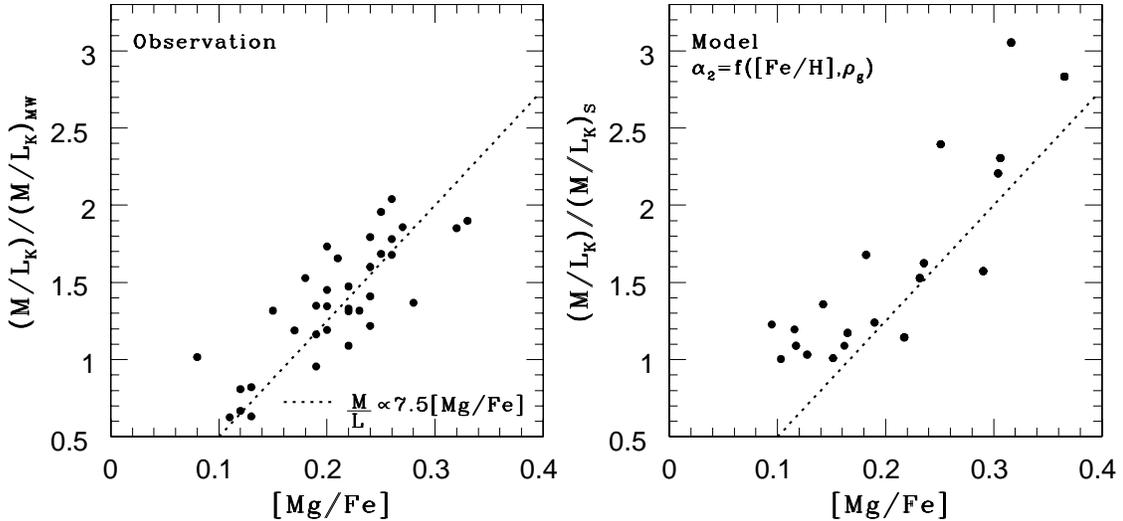}
\figcaption{
Comparison between the observed $M/L_{\rm K}-$[Mg/Fe] correlation by CV12 (left) and the predicted
one (right) for 20 models with a variable Kroupa IMF model
with $\gamma=2.0$ (i.e., $\alpha_2=f({\rm [Fe/H]},\rho_{\rm g}$)).
The observed $M/L_{\rm K}$ is normalized by the MW value ($(M/L_{\rm K})_{\rm MW}$)
whereas the predicted one is normalized by the Salpeter one for 
a solar metallicity and an age of 12.6 Gyr ($(M/L_{\rm K})_{\rm S}$)
from the MILES code 
by Vazdekis et al.  (2010).  
This normalization is done so that the simulated range of $M/L$ can be similar
to the observed one for a better comparison.
A dotted line describing
$M/L \propto 7.5 {\rm [Mg/Fe]}$ is shown for the two panels so that the observed and predicted
correlations can be better compared.
\label{fig-5}}
\end{figure*}

\begin{figure*}
\plotone{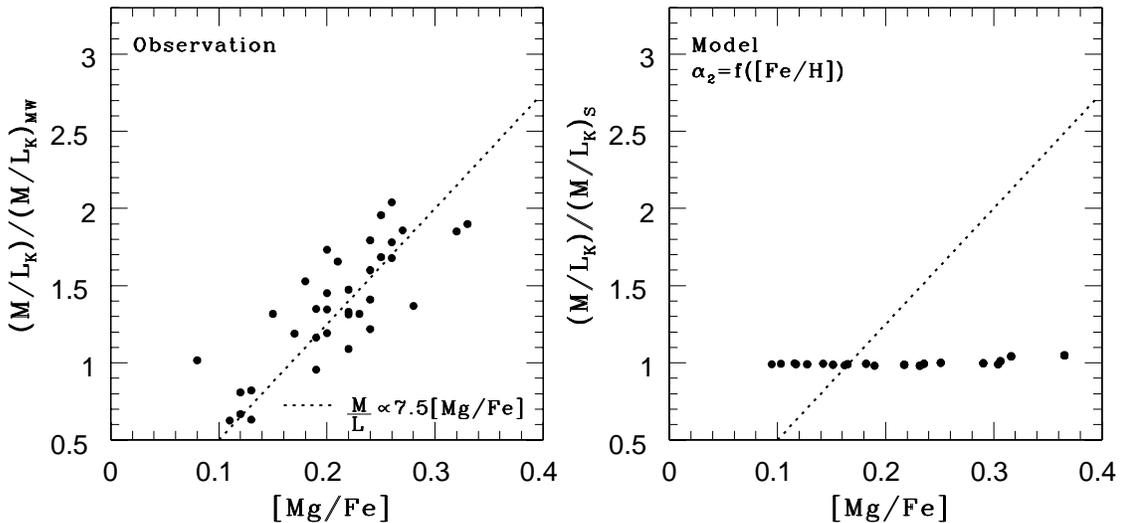}
\figcaption{
The same as Figure 5 but for the models with $\alpha_2$ depending  only on [Fe/H]
(i.e.,  $\gamma=0$;  $\alpha_2=f({\rm [Fe/H]})$).
\label{fig-6}}
\end{figure*}

\begin{figure*}
\plotone{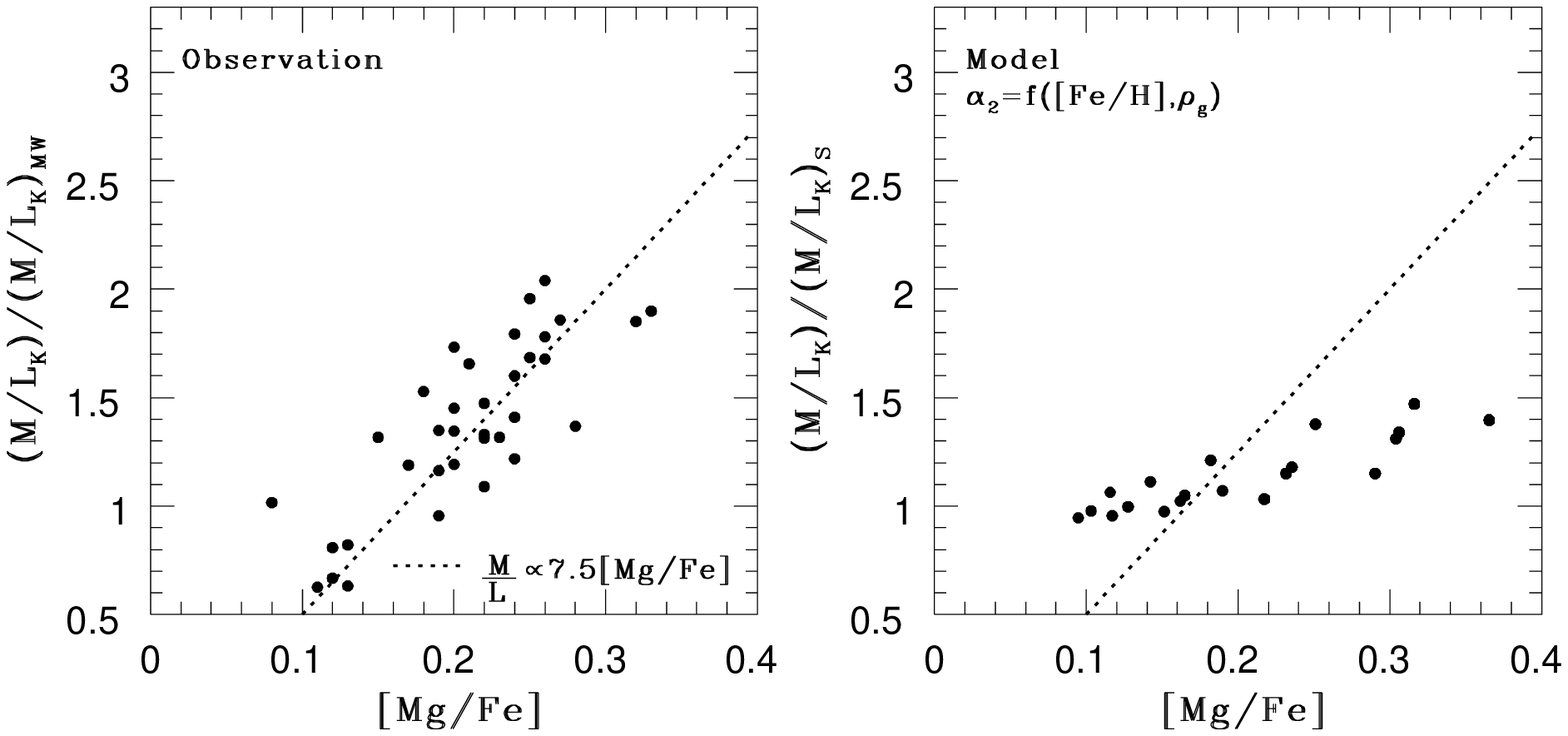}
\figcaption{
The same as Figure 5 but for the models with  
$\gamma=1.0$ (weaker dependence of $\alpha_2$ on $\rho_{\rm g}$).
\label{fig-7}}
\end{figure*}

\subsection{Comparison with observations}

Figure 5 shows a comparison between the observed $M/L_{\rm K}-$[Mg/Fe] correlation
for elliptical galaxies (CV12) 
and the  predicted one for $\gamma=2$.
If $\alpha_2$ is assumed to be  0.5[Fe/H]+$2\log \rho_{\rm g}$+2.3, 
then the predicted correlation appears to be very similar to the observed one both in
the slope and in the dispersion in $M/L_{\rm K}$. Although the ways to derive $M/L_{\rm K}$
in observations and models
are not exactly the same with each other, 
this similarity would suggest 
that $\alpha_2$ depends on $\rho_{\rm g}$  in star-forming gas clouds of elliptical
galaxies at their formation.
In the present models with $\alpha_2 \propto  \beta {\rm [Fe/H]} +\gamma \log \rho_{\rm g}$,
$\alpha_2$ and $M/L_{\rm K}$ can be both larger in galaxies with higher $\rho_{\rm g}$ for which
star formation can proceed more rapidly so that [Mg/Fe] can be higher
for $\gamma >0$. However, in order to reproduce the observed steep dependence of $M/L_{\rm K}$
on [Mg/Fe] (CV12), $\gamma$ needs to be as large as 2.
These results imply that the observed slope in the $M/L_{\rm K}-$[Mg/Fe] correlation of elliptical
galaxies can be used to give strong constraints on the IMF model for low-mass stars in
forming galaxies.

Figure 6 shows a comparison between the observed $M/L_{\rm K}-$[Mg/Fe] correlation
for elliptical  galaxies (CV12) 
and the predicted one for  $\gamma=0$ (i.e., no dependence on $\rho_{\rm g}$).
Clearly, the predicted $M/L_{\rm K}-$[Mg/Fe] correlation is qualitatively inconsistent
with the observed one, which means that [Fe/H] is not a key parameter for the observed
$M/L_{\rm K}-$[Mg/Fe] correlation. 
Given that galaxies with higher [Mg/Fe] are likely to
have lower [Fe/H] in the present chemical evolution models,
these results
mean that there can be  no/little correlation between [Fe/H] and $M/L_{\rm K}$ 
(and between [Fe/H] and $\alpha_2$).
It should be noted here that a positive [Fe/H]-$\alpha_2$ correlation is
suggested by previous observations by Cenarro et al. (2003),
though such a correlation was not found in CV12.

Figure 7 shows that the predicted $M/L_{\rm K}-$[Mg/Fe] correlation
in the models with  $\gamma=1$ is significantly shallower than
the observed one. If the dependence of $\alpha_2$ on $\rho_{\rm g}$ is weaker,
then neither the high $M/L_{\rm K}$ ($>1.5$) nor the steep slope in 
the observed $M/L_{\rm K}-$[Mg/Fe] correlation can be quantitatively reproduced in the present
models. These results confirm the importance of $\rho_{\rm g}$ in controlling the IMF slope
for low-mass stars in galaxies. 
It is confirmed that if $\gamma = 2.5$, then the final mean $\alpha_2$ in some models
can be  too large ($3.5$) to
be consistent with observations. Therefore, $\gamma$ needs to be in a certain range for
successful reproduction of observations.
Given that $P_{\rm g}$ depends on $\rho_{\rm g}$ through a thermodynamic equation,
the derived steep dependence of $\alpha_2$ on $\rho_{\rm g}$ implies that
$P_{\rm g}$ could  be also  a key parameter for the IMF slope of low-mass stars.

\begin{figure*}
\epsscale{0.8}
\plotone{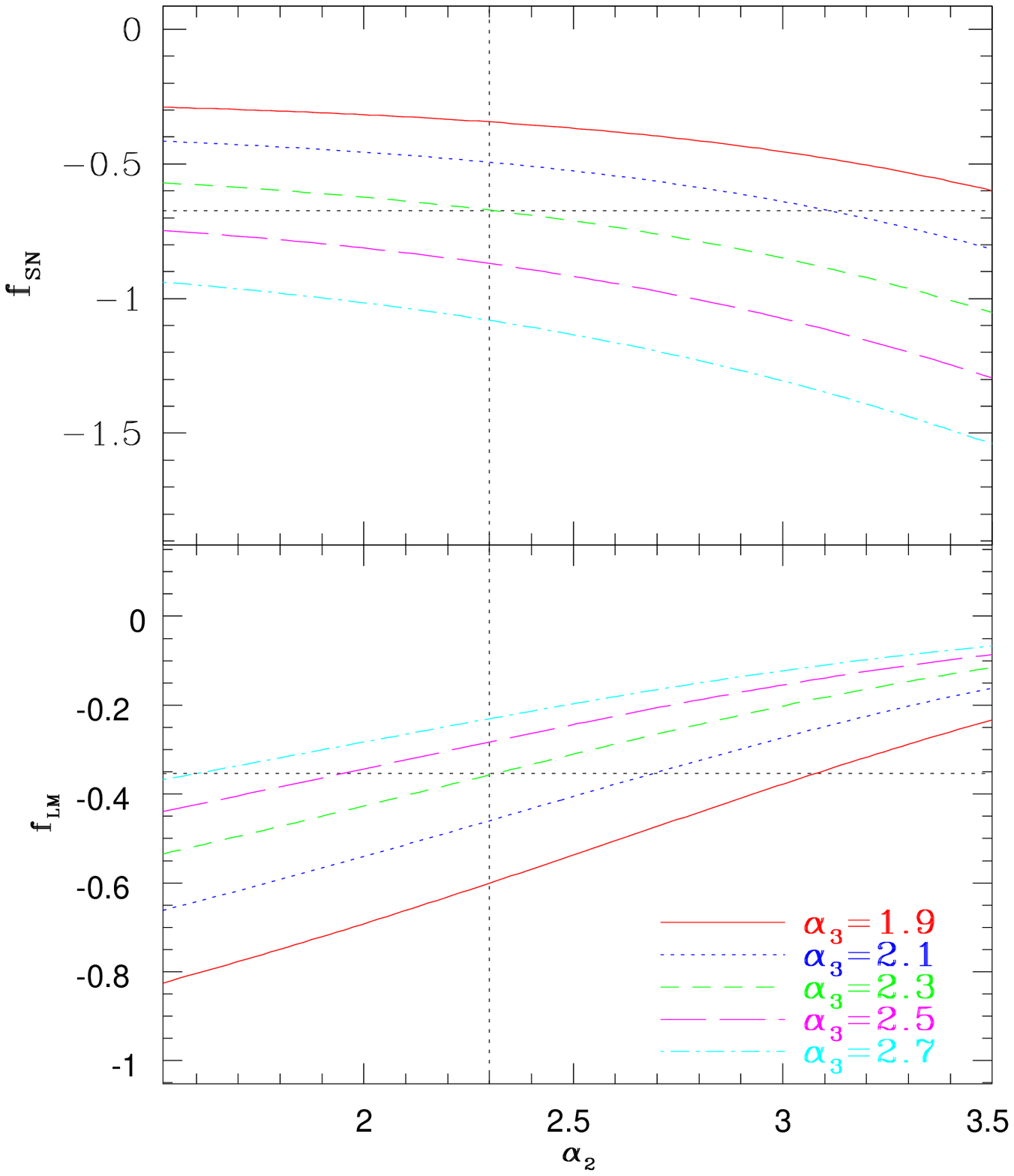}
\figcaption{
The mass fraction of high-mass stars with $m \ge 8 M_{\odot}$ ($f_{\rm SN}$, upper)
and that of low-mass stars with  $m \le 1M_{\odot}$ ($f_{\rm LM}$, lower)
as a function of $\alpha_2$ for $\alpha_3$=1.9 (red, solid), 2.1 (blue, dotted), 
2.3 (green, short-dashed), 
2.5 (magenta, long-dashed), and 2.7 (cyan, dot-dashed). The canonical $\alpha_2$ (2.3)
and $f_{\rm SN}$ and $f_{\rm LM}$ 
estimated for the canonical IMF with $\alpha_2=\alpha_3=2.3$ are shown
by dotted lines for comparison.
\label{fig-8}}
\end{figure*}

\section{Discussion}

\subsection{Origin of the observed different dependences of the IMF slope $\alpha_2$
on [Fe/H] between dwarfs and giant
elliptical galaxies}

Geha et al. (2013) have recently shown that the IMF slopes ($\alpha_2$) of dwarf
galaxies in the Local Group (e.g., LMC, SMC, and ultra-faint dwarfs) are steeper in
more metal-rich systems.
Although the total number of galaxies with known IMF slopes for low-mass stars
is still very small,
this result appears to be  consistent at least qualitatively
with the proposed IMF dependence on [Fe/H] by
M12. 
This result, however,  appears to be  inconsistent with the result by CV12
which have found no clear [Fe/H]-dependence
of the IMF slope in giant elliptical galaxies. 
It should be noted here that the [Fe/H] range of  elliptical galaxies in CV12
is only 0.3 dex and thus corresponds to only 0.15 variation in $\alpha_2$ (for the derived
relation of $\alpha_1 \propto 0.5 \times {\rm [Fe/H]}$ in Geha et al. 2013). 
The current observational data would not enable us to clearly distinguish between such a small $\alpha_2$
variation and no $\alpha_2$ variation with [Fe/H] in elliptical galaxies.
The apparently inconsistent results by CV12 and Geha et al. (2013)  could possibly mean  that 
the IMF slope for low-mass stars does not depend solely  on [Fe/H]. 
So a key question here is why dwarfs appear to show more clearly the dependence of $\alpha_2$
(or $\alpha_1$) on [Fe/H].

As shown in Figure 3 of the present study,  $\alpha_2$ depends almost linearly on 
[Fe/H] in the early chemical evolution phases 
(i.e., $-2 \le {\rm [Fe/H]} \le -0.5$) of forming elliptical galaxies, even though
$\alpha_2$ is assumed to depend both on [Fe/H] and $\rho_{\rm g}$. 
This result suggests that if star formation can be truncated by stellar winds
from massive stars and supernovae
in the early chemical evolution phases (${\rm [Fe/H]} < -0.5$) for dwarfs,
and if the truncation epochs are earlier for less massive dwarfs, 
then the dwarfs can show a correlation between $\alpha_2$ and [Fe/H] 
(and a mass-metallicity relation).  Thus, the truncation epoch of star formation
is a key parameter for the final $\alpha_2$ for dwarfs.

Figure 3 also shows that $\alpha_2$ can decrease after $\alpha_2$ takes its
peak values at higher metallicities ([Fe/H]$>-0.5$) owing to the dependences
of $\alpha_2$ on $\rho_{\rm g}$ (i.e., lower $\alpha_2$ in lower densities in
late gas-poor phases of galaxy formation).  Therefore,
$\alpha_2$ does not depend simply on [Fe/H] for 
galaxies that can continue star formation beyond [Fe/H]$\sim -0.5$.
Giant elliptical galaxies that formed with high star formation efficiencies 
and thus high metallicities accordingly do not show a strong dependence of $\alpha_2$
on [Fe/H]. The star formation time scale, which depends primarily on $\rho_{\rm g}$,
can be a key parameter for these metal-rich elliptical galaxies.
This is one of possible explanations  for the observed different dependences of $\alpha_2$
on [Fe/H] between dwarfs and giant elliptical galaxies. We need to investigate whether 
$\alpha_2$ depends differently on [Fe/H] between dwarfs and giant elliptical galaxies by using
self-consistent chemodynamical simulations of galaxy formation with 
a same variable IMF model in our future study.

\subsection{Bottom-heavy $\alpha_2$ yet slightly top-heavy $\alpha_3$ ?}

Greggio \& Renzini (2012, GR12) convincingly discussed whether top-heavy/bottom-heavy IMFs
can really explain a number of key  observed properties of elliptical galaxies
{\it in a self-consistent manner}. They clearly demonstrated that the canonical IMF
($\alpha=2.3$) can better explain both (i) $B$-band mass-to-right ratios and their correlations
with mean stellar ages observed in elliptical galaxies
and (ii) oxygen and silicon mass-to-light ratios 
($M_{\rm O}/L_{\rm B}$ and $M_{\rm Si}/L_{\rm B}$, respectively) in clusters
of galaxies. They furthermore showed that both bottom-heavy and top-heavy IMFs
fail to explain these observations by using some idealized models of elliptical galaxy
formation. Then, how can we explain these two observations, if elliptical galaxies really
have a  bottom-heavy IMF for low-mass stars ?
We provide a clue to this puzzling problem regarding the IMF of elliptical galaxies as follows.

Since GR12 adopted a simple IMF model with a single power-law slope in discussing the above 
IMF problems,
a bottom-heavy IMF means bottom-heavy for both low-mass and high-mass
stars. On the other hand,  $\alpha_2$ for low-mass stars and $\alpha_3$ for high-mass stars
can vary independently from each other in a  variable Kroupa IMF model
so that a bottom-heavy IMF for low-mass stars does not
necessarily mean bottom-heavy for high-mass stars. As a result of this, elliptical galaxies
with bottom-heavy IMFs for low-mass stars can have as high  
$M_{\rm O}/L_{\rm B}$ and $M_{\rm Si}/L_{\rm B}$
as the canonical IMF predicts, if $\alpha_3$ is only slightly top-heavy.
Figure 8 shows the mass fraction of stars with $m \ge 8 M_{\odot}$ ($f_{\rm SN}$) as a function
of $\alpha_2$ for five $\alpha_3$ values.  This $f_{\rm SN}$ can be used as a
more accurate measure for
$M_{\rm O}/L_{\rm B}$ and $M_{\rm Si}/L_{\rm B}$, because oxygen and silicon abundances come 
largely from Type II SNe (GR12). In this simple model,
$f_{\rm SN}$ is 0.21 for the canonical IMF with $\alpha_2=\alpha_3=2.3$ and
$f_{\rm SN}$ at a given $\alpha_2$ is lower for larger $\alpha_3$ (i.e., more bottom-heavy). However,
$f_{\rm SN}$ can be as high as 0.21 for $\alpha_3=2.1$
even for  $\alpha_2 \sim 3$.
This implies  that the observed bottom-heavy IMF for 
low-mass stars is not inconsistent with the observed
$M_{\rm O}/L_{\rm B}$ and $M_{\rm Si}/L_{\rm B}$, as long as the IMF for high-mass stars
is only slightly top-heavy.

Figure 8 also shows the mass fraction of low-mass stars with  $m \le 1M_{\odot}$ 
($f_{\rm LM}$) as a function of $\alpha_2$ for five $\alpha_3$ values.
If a galaxy is dominated by low-mass  stars (i.e., higher $f_{\rm LM}$),
then the galaxy can have rather high  $M/L_{\rm B}$ (GR12).
It is clear that the larger $\alpha_2$ is, the higher $f_{\rm LM}$ is (i.e., more dwarf-dominated),
independent of $\alpha_3$.  The model with a slightly top-heavy IMF for high-mass stars
($\alpha =1.9$) can show $f_{\rm LM}$ at $\alpha_3=3$ as low as  $\sim 0.44$  estimated
for the canonical IMF with $\alpha_2=\alpha_3=2.3$. This result means
that even if the IMF for low-mass stars is bottom-heavy,  $M/L_{\rm B}$ can not 
be so high (i.e., can not be dominated by low-mass dwarf stars), as long as the IMF for high-mass stars
is slightly top-heavy. This result suggests that the observed range of $M/L_{\rm B}$
($2-14$ shown in GR12) in elliptical galaxies is not inconsistent with the observed bottom-heavy
IMF for low-mass stars.
Thus it  is possible that future models of elliptical galaxy formation
with a variable Kroupa IMF can explain a number of observational results regarding
the IMF slopes in a fully self-consistent manner.

Although elliptical galaxies with bottom-heavy IMF for low-mass stars and slightly
top-heavy IMF in high-mass stars might form, as the present study suggests,
it is theoretically unclear why such a combination of bottom-heavy/top-heavy IMFs
is possible in a single star-forming gas cloud.
Elmegreen (2004) considered that different parts of the IMF can be independently
determined and thereby demonstrated that the entire IMF can be constructed by
using three log-normals, each of which has its own characteristic stellar mass.
The shape of his multi-component IMF model 
depends basically on the amplitudes of the three log-normals: It would be possible in principle
that a certain combination of the three amplitudes can yield an IMF with bottom-heavy $\alpha_2$
and top-heavy $\alpha_3$.
If we understand how the basic nine parameters determining the amplitudes of the three log-normals
depend on physical properties of star-forming cloud, such as [Fe/H], interstellar radiation fields,
$\rho_{\rm g}$, 
and $P_{\rm g}$,
then we could better understand in what physical conditions of a star-forming cloud
an IMF with a bottom-heavy $\alpha_2$ and a top-heavy $\alpha_3$ is possible.
Thus extensive investigation on the dependences of the basic parameters of the IMF
on physical properties of star-forming gas clouds will greatly advance our understanding
of the origin of the IMF in elliptical galaxies.

\section{Conclusions}

We have investigated the origin of the observed bottom-heavy IMF in elliptical galaxies
by using one-zone chemical evolution models.
A principal assumption is that the Kroupa IMF slope $\alpha_2$
depends both on metallicities ([Fe/H]) and gas densities ($\rho_{\rm g}$) of star-forming
gas clouds in such a way that $\alpha_2$ is proportional to 
$\beta {\rm [Fe/H]} + \gamma \log \rho_{\rm g}$ ($\beta$ is fixed at 0.5).
We have searched for the best parameter value
of $\gamma$ that can reproduce the observed $M/L_{\rm K}-{\rm  [Mg/Fe]}$ relation
(corresponding to $\alpha_2-{\rm [Mg/Fe]}$ relation) in elliptical galaxies.
Although the present model for $M/L$ has some limitations (e.g., using SSPs for a fixed  IMF),
we have found the following important results.

(1) Our chemical evolution models with $\alpha_2 = 2.3+0.5{\rm [Fe/H]} + 2 \log \rho_{\rm g}$
(i.e., $\gamma=2$) 
can reproduce the observed positive $M/L_{\rm K}-{\rm  [Mg/Fe]}$ correlation
(i.e., higher $M/L_{\rm K}$ for higher [Mg/Fe]) in a quantitative manner.
Furthermore, some  models with $\gamma=2$ can show larger $\alpha_2$ ($\sim 3$),
which is consistent with recent observations (e.g., CV12). 
However, our models with low $\gamma$ (=0 and 1) can not reproduce the 
$M/L_{\rm K}-{\rm  [Mg/Fe]}$ correlation. These results suggest that the IMF slope for low-mass
stars needs to  depend more strongly on $\rho_{\rm g}$ than on [Fe/H]. 

(2) The observed different dependences of $\alpha_2$ on [Fe/H] in dwarf and giant
elliptical galaxies can not be simply explained by a variable IMF model that depends
only [Fe/H] for low-mass stars. Instead, such differences suggest that $\alpha_2$
needs to depend both on [Fe/H] and $\rho_{\rm g}$. A key parameter for $\alpha_2$
is suggested to be the truncation epoch of star formation for dwarfs and the time
scale of star formation for giant elliptical galaxies.
It is our future study to understand why gas density can be  a fundamental parameter
for $\alpha_2$.

(3) The observed bottom-heavy IMF for low-mass stars ($\alpha_2$)
 in elliptical galaxies would not be a problem in 
explaining the observed $M_{\rm O}/L_{\rm B}$ and $M_{\rm Si}/L_{\rm B}$
in galaxy clusters, as long as the variable Kroupa IMF 
for high-mass stars (i.e., $\alpha_3$)
is slightly more top-heavy. Both
the observed bottom-heavy IMF and the cluster metal content (for which the Salpeter IMF
is suggested to be required) could  be self-consistently
explained in a model in which $\alpha_2$ and $\alpha_3$ 
can vary independently from each other.

(4) In the variable Kroupa IMF,  the mass fraction of low-mass stars with $m\le 1M_{\odot}$ 
($f_{\rm LM}$)
depends on $\alpha_3$ for a given $\alpha_2$ such that $f_{\rm LM}$ can be lower for
smaller $\alpha_3$. Therefore,  elliptical galaxies with  $\alpha_2 \sim 3$ 
(i.e., bottom-heavy) can have $f_{\rm LM}$ as low as 0.44 estimated for the canonical
IMF with $\alpha_2=\alpha_3=2.3$, 
if the IMF for high-mass stars is only slightly top-heavy ($\alpha_3 \sim 2$).
This implies that the observed  $B$-band mass-to-light ratios ($M/L_{\rm B}$) in elliptical galaxies
are not inconsistent with a bottom-heavy IMF for low-mass stars. A more detailed modeling
with variable $\alpha_2$ and $\alpha_3$ is necessary to confirm that both the bottom-heavy $\alpha_2$
and $M/L_{\rm B}$ in elliptical galaxies can be self-consistently explained by a variable IMF
model.

The present study suggests that the Kroupa IMF slopes, $\alpha_2$
and $\alpha_3$, would need to vary
independently from each other for more self-consistent explanations of different observational
results regarding possible IMF variations in elliptical galaxies.
 Since the present study did not extensively discuss the physics behind this independently
varying IMF slopes, it is our future study to investigate why and how the three IMF slopes
depend on physical properties of star-forming gas clouds.
Although the present study has adopted somewhat idealized models for elliptical galaxies formation,
the formation processes are significantly more complicated in
recent hierarchal galaxy formation models (e.g., Naab 2012).
Thus, it is our future study to investigate whether the variable Kroupa IMF model can really explain 
the observed bottom-heavy IMF of elliptical
galaxies in a  more sophisticated formation model of elliptical galaxies.

\acknowledgments
I (K. B.) am grateful to the anonymous referee for constructive and
useful comments.
K.B. acknowledges the financial support of the Australian Research
Council throughout the course of this work.

\appendix

\section{[Mg/Fe]$-$[Fe/H] relations for different Kroupa  IMFs}

In order to demonstrate that final [Mg/Fe] in a 
one-zone chemical evolution model does not depend
so strongly on $\alpha_1$ and $\alpha_2$ of the (fixed) Kroupa IMF,
we have investigated the models with different $\alpha_1$ and $\alpha_2$
(yet a fixed $\alpha_3=2.3$). These models (M21$-$M24) have $t_{\rm a}=0.5$ Gyr,
$t_{\rm trun}/t_{\rm a}=2$, and $C_{\rm sf}=0.4$ so that they can be compared
with the model M1 with the Salpeter IMF. Figure 9 shows that the final [Mg/Fe]
is very similar between these models with different $\alpha_1$ and $\alpha_2$
and M1 with the Salpeter IMF,
though final [Fe/H] depends on these IMF slopes. The present study adopted an approximation
of a fixed IMF slope ($\alpha_3=2.35$, i.e., Salpeter IMF) in chemical evolution
models,  though $\alpha_1$ and $\alpha_2$
are assumed to vary with time at each time step in the models.
The results in Figure  9 mean that if the time evolution of chemical yields
due to varying $\alpha_1$ and $\alpha_2$ is included in one-zone models,
the present results can not change significantly.
These results therefore  demonstrate that the adopted  approximation of
a fixed IMF slope in one-zone models
can be justified and thus 
good enough to discuss the final [Mg/Fe] of elliptical galaxy formation
models in the present study.

\section{Possible $M/L$ differences between variable Kroupa and single-power-law
IMFs}

In order to discuss how $M/L$ could be possibly different between variable Kroupa
and single-power-law IMFs, we have investigated (i) the mass fraction ($f_{\rm SN}$) 
 of massive
stars with stellar masses ($m$) equal to or larger than $8 M_{\odot}$ (i.e., those
which explode as SNII and leave stellar remnants) and the mass fraction
($f_{\rm LM}$) of low-mass stars with $m \le 1 {\rm M}_{\odot}$.  Since we can 
not directly estimate $M/L$, we discuss the possible $M/L$ differences between
variable Kroupa and single-power-law IMFs by using these $f_{\rm SN}$ and $f_{\rm LM}$. 
By definition, a variable Kroupa IMF has $\alpha_1=\alpha_2-1$ and $\alpha_3=\alpha_2$
whereas  a single-power-law IMF has $\alpha_1=\alpha_2=\alpha_3$.
A galaxy can have a larger number of stellar remnants for a more top-heavy IMF 
so that $f_{\rm SN}$ (thus $M/L$)  can be larger.
A galaxy is more dominated by dwarf stars for a more bottom-heavy IMF so that $f_{\rm LM}$
(thus $M/L$) can be larger.

Figure  10  shows that (i) $f_{\rm SN}$ is slightly  larger in variable Kroupa IMFs
than in single-power-law IMFs, and (ii) $f_{\rm LM}$ is larger in
single-power-law IMFs than in variable Kroupa IMFs, and (iii)
$f_{\rm SN}$ differences between the two IMF models are significantly smaller
in comparison with $f_{\rm LM}$ differences.
For example, $f_{\rm SN}$ ($f_{\rm LM}$) is 0.21 (0.45) for a variable Kroupa IMF with
$\alpha_2=2.3$ and 0.16 (0.58) for a single-power-law IMF with $\alpha_2=2.3$.
These results therefore mean that $M/L$ should be systematically larger in 
single-power-law IMFs than in variable Kroupa IMFs for a given $\alpha_2$ ($=\alpha_3$).
The bottom line in this figure is that
$f_{\rm LM}$ is always slightly larger in single-power-law IMFs than in the variable Kroupa IMFs
and the $f_{\rm LM}$ differences do not depend strongly  on $\alpha_2$.

These results indicate  that $M/L$ can be always slightly  overestimated
by a very similar amount in the present different models with different $\alpha_2$
(in comparison with the true $M/L$ for variable Kroupa IMFs).  These therefore
demonstrate  that the slope in the simulated $M/L-{\rm [Mg/Fe]}$ relation can be very
close to the true value (estimated self-consistently 
by using a stellar population synthesis code
for a variable Kroupa IMF). 
Accordingly, as long as we discuss the slope of the observed $M/L-{\rm [Mg/Fe]}$ relation
(i.e., not the absolute value of $M/L$ itself),
the usage of the MILES code for variable single-power-law IMFs can be justified.
However, the simulated
absolute magnitudes of $M/L$ can slightly deviate from the true ones  in the present study
so that a direct comparison between the observed and simulated absolute values of  $M/L$ 
can not be so valid.

\clearpage

\begin{figure}
\epsscale{1.0}
\plotone{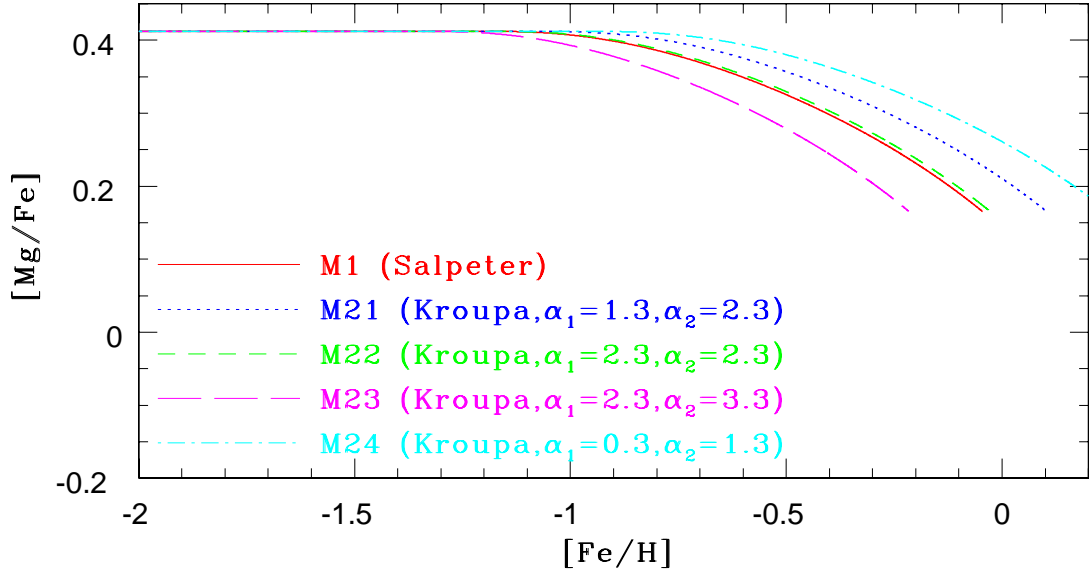}
\figcaption{
The evolution of galaxies on the [Mg/Fe]$-$[Fe/H] plane 
for the Salpeter IMF model M1 with $\alpha=2.35$ (red, solid), 
four  variable Kroupa IMF models,
M21 with $\alpha_1=1.3$ and $\alpha_2=2.3$  (blue, dotted),
M22 with $\alpha_1=2.3$ and $\alpha_2=2.3$ (green, short-dashed),
M23 with $\alpha_1=2.3$ and $\alpha_2=3.3$ (magenta, long-dashed), 
and M24 $\alpha_1=0.3$ and $\alpha_2=1.3$ with (cyan, dot-dashed).
For M21$-$M24, $\alpha_3$ is fixed at 2.3.
The model parameter for star formation histories
($t_{\rm a}$ and $t_{\rm trun}$) are exactly the same between
these five models.
\label{fig-9}}
\end{figure}

\begin{figure}
\epsscale{0.8}
\plotone{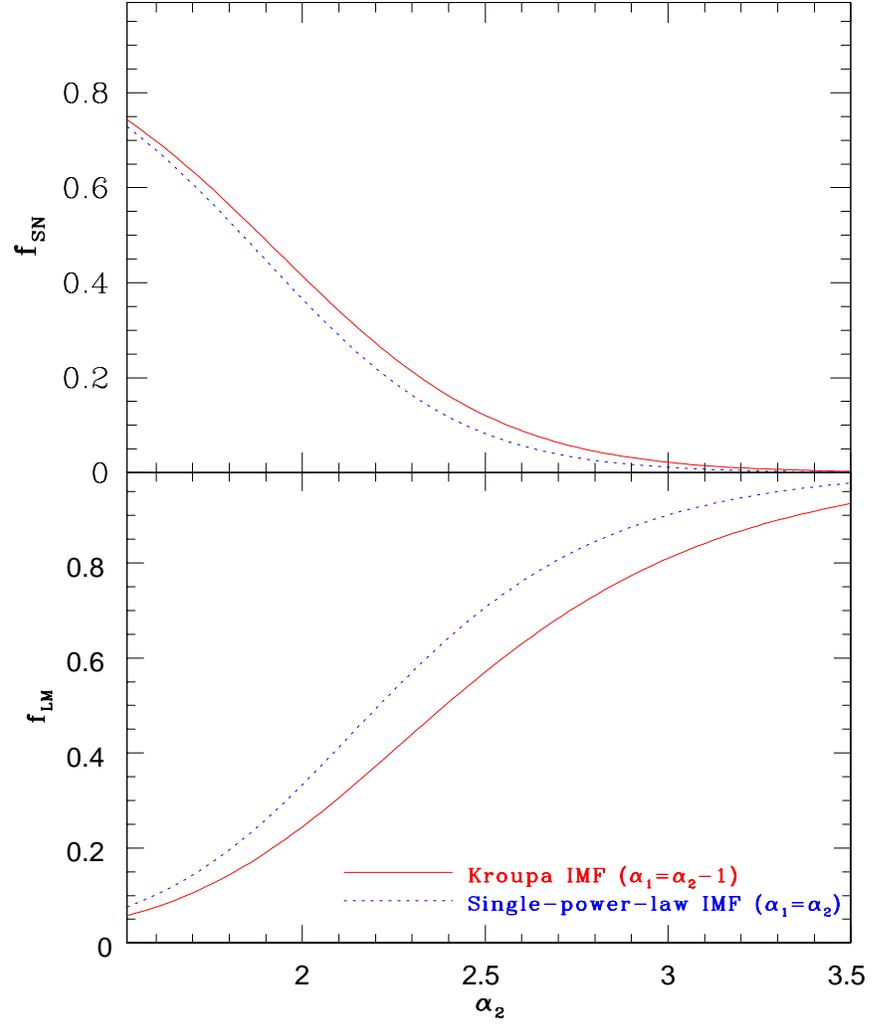}
\figcaption{
The mass fraction of high-mass stars with $m \ge 8 M_{\odot}$ ($f_{\rm SN}$, upper)
and that of low-mass stars with  $m \le 1M_{\odot}$ ($f_{\rm LM}$, lower)
as a function of $\alpha_2$ for a variable Kroupa IMF  (red, solid)
and a variable single-power-law IMF  (blue, dotted). Here $\alpha_2=\alpha_3$ is adopted.
\label{fig-10}}
\end{figure}

\end{document}